\documentclass[11pt]{IEEEtran}

\usepackage[utf8]{inputenc} 
\usepackage[T1]{fontenc}
\usepackage{url}
\usepackage{ifthen}
\usepackage{cite}

 \usepackage[dvipdfmx]{graphicx}
 \usepackage{grffile}
 \usepackage{algorithm}
 \usepackage{algpseudocode}
\makeatletter
\newcommand{\removelatexerror}{\let\@latex@error\@gobble}
\makeatother
\algnewcommand\algorithmicinput{\textbf{Input:}}
\algnewcommand\INPUT{\item[\algorithmicinput]}
\algnewcommand\algorithmicoutput{\textbf{Output:}}
\algnewcommand\OUTPUT{\item[\algorithmicoutput]}

\usepackage{color}

\usepackage{amssymb}
\usepackage[cmex10]{amsmath} 

\allowdisplaybreaks[1]

\newcommand{\F}[0]{{\mathbb F}}

\begin{document}
\title{Asymptotic Analysis of Spatial Coupling Coding for {Compute-and-Forward Relaying}\thanks{Parts of this paper {were} presented at ISIT2018.}}

\author{Satoshi~Takabe~\IEEEmembership{Member,~IEEE},
Tadashi Wadayama~\IEEEmembership{Member,~IEEE},
\thanks{The first and second authors are with the
Department of Computer Science, Faculty of Engineering, Nagoya Institute of Technology, Japan,
e-mail: s\_takabe@nitech.ac.jp, wadayama@nitech.ac.jp.}
and Masahito~Hayashi~\IEEEmembership{Fellow,~IEEE}
\thanks{The third author is with the Graduate School of Mathematics, Nagoya University, Japan. He is also affilitated with the
Shenzhen Institute for Quantum Science and Engineering, Southern University of Science and Technology, Shenzhen, China and
the Centre for Quantum Technologies, National University of Singapore, Singapore, e-mail:masahito@math.nagoya-u.ac.jp}
}

\maketitle

\begin{abstract}
Compute-and-forward (CAF) relaying is effective to increase bandwidth efficiency 
of wireless two-way relay channels.
In a CAF scheme, a relay is designed to decode a linear combination 
composed of transmitted messages from other terminals or relays.
Design for error-correcting codes and its decoding algorithms 
suitable for CAF relaying schemes remain as an important issue to be studied.
As described in this paper, we will present an asymptotic performance analysis of LDPC codes 
over two-way relay channels based on density evolution (DE).  
Because of the asymmetric characteristics of the channel, 
we use the population dynamics DE combined with DE formulas for asymmetric channels to obtain BP thresholds.
Additionally, we also evaluate the asymptotic performance of spatially coupled LDPC codes for two-way relay channels.
The results indicate that the spatial coupling codes yield improvements in the BP threshold compared with 
corresponding uncoupled codes for two-way relay channels.
{Finally, we will compare the mutual information rate and rate achievability between the CAF scheme and the MAC separation decoding scheme.
We demonstrate the possibility that the CAF scheme has higher reliability in the high-rate region.
}
\end{abstract}


\section{Introduction}
Relays with appropriate signal processing and decoding are ubiquitous in wireless communications
such as satellite communications, mobile wireless communications, and wireless local area networks.
Increasing demand for band width efficiency in wireless communications
promotes the spread of research activities on relaying and forwarding techniques.
For example, theoretical limits of efficiencies of relaying techniques 
such as decode-and-forward \cite{Cover79} and amplify-and-forward \cite{Laneman} have been investigated intensively.
Recently, Nazar and Gastpar presented a novel concept of {\em compute-and-forward (CAF) scheme} \cite{Nazer11}.
In a CAF scheme, a relay is designed to decode a linear combination 
composed of messages transmitted from other terminals (or relays).
Then the relay forwards a decoded linear combination to 
another relay or a terminal. 
That is, the repeater has no intention to decode each message separately. 
The concept is also designated as wireless network coding 
or {\em physical layer network coding},
 which has attracted strong research interest \cite{Katti08, Zhang09}.
Recently, Sula \cite{Sula17} {et al.} presented a practical decoding scheme for LDPC codes 
in compute-forward multiple access (CFMA) systems.
Ullah et al. \cite{Ullah17} derived the random coding error exponent for 
the uplink phase of a two-way relay channel.

\begin{figure}[!t]
\centering
\includegraphics[width=0.8\linewidth]{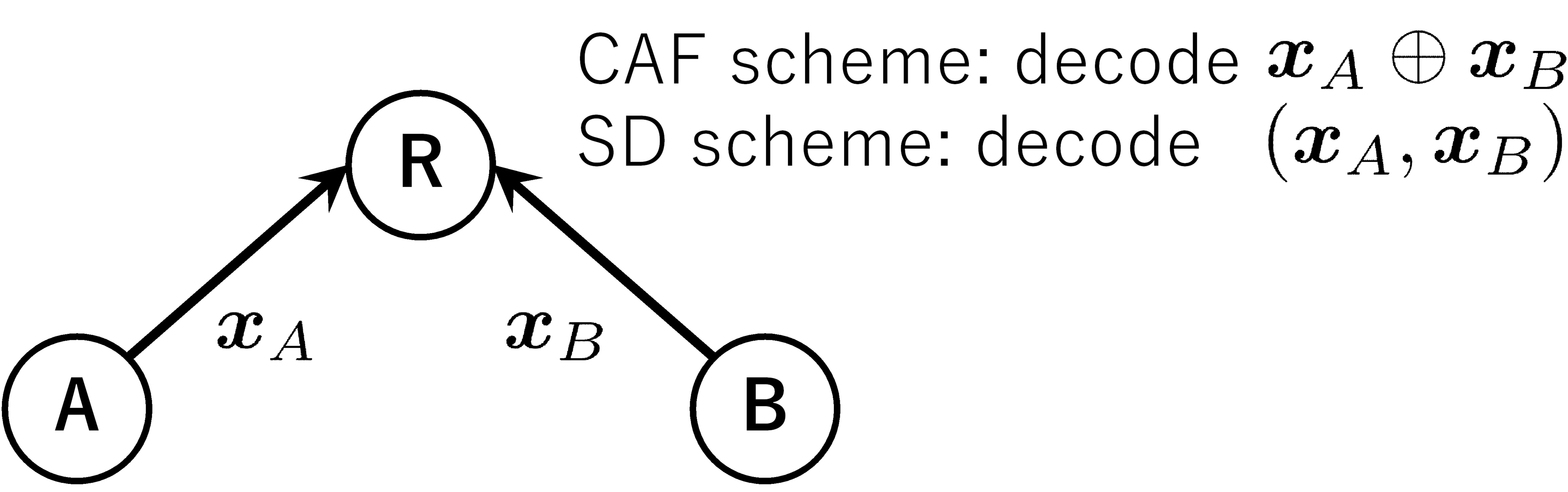}
\caption{MAC phase of the compute-and-forward (CAF) scheme and the separation decoding (SD) scheme on a two-way relay channel.}\label{zu_0}
\end{figure}

The simplest scenario for a CAF scheme might be 
{\em wireless two-way relay channels} \cite{Narayanan}.
Two terminals A and B and a relay R are involved in this channel.
Terminal A has its own message and is programmed to send 
it to terminal B. Similarly, terminal B is programmed to send 
its own message to A.
No direct wireless connection exists between A and B, but a relay R has 
bi-directional wireless connections to both of A and B.
{In the multiple-access (MAC) phase, two terminals send their messages to relay R. Then 
R attempts to decode the linear combination of their messages {as shown in Fig.~\ref{zu_0}}.
In the broadcasting phase, the decoded linear combination of messages is sent back to two terminals.}
Terminals A and B can recover an intended message 
by subtracting its own message from the received message.

{To establish a highly reliable CAF scheme, we must obtain a reliable estimate of linear combination at the relay in the MAC phase,
whereas conventional coding techniques for simple communication channels are available in the broadcasting phase.
In the MAC phase, }
appropriate error-correcting codes should be exploited because
the received signal is distorted by additive noise.
In such a case, the relay R intends to {\em decode} a sum of two codewords 
sent from A and B. One candidate of error correcting codes 
for such a situation is {\em low-density parity-check (LDPC) codes} \cite{Gallager63}.
A combination of LDPC codes and belief propagation (BP) decoding 
has been demonstrated as very powerful and effective techniques for additive noise channels \cite{MacKay99}.
Sula et al.~\cite{Sula17} discussed an appropriate modified BP decoding for 
two-way relay channels. They presented a performance analysis of 
LDPC codes over a two-way relay channel based on computer simulations.

The goals of this work are {three}. The first goal is to provide 
an asymptotic performance analysis for LDPC codes over two-way relay channels
based on {\em density evolution} (DE). DE \cite{Richardson} is a common theoretical tool 
to study the asymptotic typical behavior of a BP decoder. It provides BP thresholds of the
target channel. Although the BP threshold is below the Shannon limit, a BP threshold 
denotes a practical achievable rate with low complexity encoding and decoding.
One technical challenge for evaluating the BP threshold of two-way relay channels comes
from an asymmetric  characteristics of the channel: we cannot rely on the zero codeword 
assumption commonly used in DE analysis for binary-input memoryless output-symmetric channels \cite{Richardson}.
To overcome this difficulty, we will employ population dynamics DE \cite{Mezard}
combined with the DE formula derived by Wang et al. for asymmetric channels \cite{Wang05}.

The second goal of these assessments is to provide DE analysis for spatially coupled LDPC (SC-LDPC) codes over two-way relay channels.
It is known that appropriately designed spatially coupled codes yield improvements in BP thresholds compared with 
those of uncoupled regular LDPC  codes with comparable parameters \cite{Zigangirov99} \cite{Lentmaier10}.
In many cases, we can observe {\em threshold saturation} \cite{Kudekar13}, i.e., a phenomenon 
by which the BP threshold converges to the MAP threshold.
The same is true for spatially coupling coding for two-way erasure multiple access channels
for a joint CAF scheme~\cite{Hern}.
Typical behavior of BP decoding of 
spatially coupled LDPC codes over 
the two-way relay channels is unknown, except for erasures. 
We consider the study as worthwhile, not only from practical interests but also from theoretical interests,
to provide an example of the DE analysis for general asymmetric channels.
In this work, we extend
the population dynamics DE to protograph codes~\cite{Thorpe04} and  
perform numerical evaluations 
{in a similar way in~\cite{Hassani} for spatially coupled constraint satisfaction problems.}

The third goal of this paper is comparison between 
the CAF scheme and the MAC {separation decoding (SD)} scheme.
{In the SD scheme for the MAC phase of two-way relay channel,
  relay R attempts to decode a pair of messages from two terminals separately~\cite{Yedla09} as shown in Fig.~\ref{zu_0}.
 Although it is a natural and conventional approach, it is sometimes sub-optimal, as shown in~\cite{Nazer07} in the case of binary symmetric channels.}
We first compare both schemes {in our setting}
when we use random coding and maximum likelihood decoding.
In the next step, we compare 
our obtained LDPC codes with LDPC codes of the SD scheme.
For this comparison, we recall the numerical analysis for the {BP} threshold reported in \cite{Yedla}.

{This paper is organized as follows.
In Sec.~\ref{S2}, we introduce problem settings of LDPC coding and BP decoding on two-way relay channels. 
In Sec.~\ref{S3}, we describe the population dynamics DE for two-way relay channels.
Because of its asymmetric nature, we combine the DE formulas for asymmetric channels
 with the population dynamics algorithm for numerical evaluation.
We then extend it to the case of SC-LDPC codes to assess its typical decoding performance.
In Sec.~\ref{S4}, we compare the mutual information and rate achievability between the CAF scheme and the SD scheme.
The last section is devoted to a summary and discussions.
}

\section{Preliminaries}\label{S2}

\subsection{Problem setting}\label{S2-1}

The wireless channel model assumed here is described as shown below.
Let $X_A^{(t)}$ (resp. $X_B^{(t)}$) be a binary random variable where
$t = 1,2, \ldots$ represents a time index.
The binary--bipolar 
conversion function $\mu: \{0,1\} \rightarrow \{+1, -1\}$,
$\mu(x) \triangleq 1 - 2 x$ is applied to $X_A^{(t)}$ and $X_B^{(t)}$ 
before their transmission. Therefore, we assume 
binary phase shift keying (BPSK) as a modulation format.
Terminals A and B then transmit the modulated signals 
$\mu(X_A^{(t)})$ and $\mu(X_B^{(t)})$ to the air. The relay R observes 
a received symbol 
\begin{equation} \label{channel_model}
	Y^{(t)} = \mu(X_A^{(t)}) + \mu(X_B^{(t)}) + W^{(t)},
\end{equation}
where $W^{(t)}$ is a zero mean Gaussian random variable with variance $\sigma^2$.
The channel model (\ref{channel_model}) is justified 
under the assumption such that 
perfect symbol/phase synchronization and perfect power control 
are achieved at R.
The relay R is designed to infer $X_A^{(t)} \oplus X_B^{(t)}$ 
 from $Y^{(t)}$ 
as correctly as possible, where the operator $\oplus$ represents the addition over $\F_2$.  

If no error-correcting code is used, then symbol by symbol estimation is applicable.
In the next phase, 
the estimate $\hat x_A^{(t)} \oplus \hat x_B^{(t)}$
is then broadcasted to A and B. 
If the estimate $\hat x_A^{(t)} \oplus \hat x_B^{(t)}$ equals 
the true value $x_A^{(t)} \oplus x_B^{(t)}$, then 
terminal A (resp. B) can retrieve $x_B^{(t)}$ (resp. $x_A^{(t)}$) 
from $\hat x_A^{(t)} \oplus \hat x_B^{(t)}$.
This protocol 
can be regarded as the simplest case of the CAF technique \cite{Nazer11}. It 
increases the bandwidth efficiency of the {two-way relay channel}.

\subsection{LDPC coding}\label{S2-2}

{As described in this paper, we restrict ourselves to the case in which two terminals use the same LDPC codes $C \subset \F_2^n$.}
Terminals A and B independently select their own codewords 
$\mathbf{x}_A  \triangleq (x_{A,1},\ldots, x_{A, N})  \in C$
and $\mathbf{x}_B  \triangleq (x_{B,1},\ldots, x_{B, N}) \in C$ 
according to their own message. From the channel model 
(\ref{channel_model}), the received word is given as
\begin{equation}
	\mathbf{y} 
	\!=\!  (\mu(x_{A,1}),\dots, \mu(x_{A, N})) + (\mu(x_{B,1}),\dots, \mu(x_{B, N})) 
	+ \mathbf{w},
\end{equation}
where $\mathbf{w}$ represents additive white Gaussian noise vector.
A decoder, possibly a BP decoder, is programmed to recover 
$\mathbf{x}_A \oplus \mathbf{x}_B$ from the received word $\mathbf{y}$.
As described herein, we specifically examine decoding methods for recovering $\mathbf{x}_A \oplus \mathbf{x}_B$.

\subsection{Degraded channel}\label{S2-3}

Assume that two stochastic processes {$\{X_A^{(t)}\}$ and $\{X_B^{(t)}\}$}
are IID and that $X_A^{(t)}$ and $X_B^{(t)}$ are independent.
For simplicity, we assume that $\mathrm{Pr}[X_A^{(t)} = 1] = \mathrm{Pr}[X_B^{(t)} = 1] = 1/2$ holds for 
any $t$. From these assumptions, we have the following probability of events:
\begin{eqnarray}
\mathrm{Pr} [\mu(X_A^{(t)}) + \mu(X_B^{(t)}) = 0] &=& 1/2,	\\
\mathrm{Pr} [\mu(X_A^{(t)}) + \mu(X_B^{(t)}) = -2] &=& 1/4,	\\
\mathrm{Pr} [\mu(X_A^{(t)}) + \mu(X_B^{(t)}) = 2] &=& 1/4.	
\end{eqnarray}
Let $Z^{(t)} = X_A^{(t)} \oplus X_B^{(t)}$. From the IID assumption,
$Z^{(t)}$ is also a memoryless stochastic process.
We now consider a virtual channel called {{\em degraded channel}} with 
input and output symbols respectively denoted as $Z^{(t)}$ and $Y^{(t)}$.
It is evident that the prior probability of $Z^{(t)}$ is given as $\mathrm{Pr}(Z^{(t)} = 0) = \mathrm{Pr}(Z^{(t)} = 1) = 1/2$.
The conditional PDF representing the channel statistics 
of the {degraded channel} is then given as
\begin{equation} 
\begin{aligned}
\!\mathrm{Pr} [Y^{(t)} \!&=\! y | Z^{(t)} \!= 1] \!=\! F(y; 0, \sigma^2), \\ 
\!\mathrm{Pr} [Y^{(t)} \!&=\! y | Z^{(t)} \!= 0] \!=\! \frac{1}{2} F(y; -2, \sigma^2) \!+\!\frac{1}{2} F(y; 2, \sigma^2), 
\end{aligned}\label{vir}
\end{equation}
where $F(y; m, \sigma^2)$ is the Gaussian distribution with mean $m$ and variance $\sigma^2$
defined as
\[
F(y; m, \sigma^2)  \triangleq \frac{1}{\sqrt{2 \pi \sigma^2}} \exp \left( \frac{-(y-m)^2}{2 \sigma^2} \right).
\]

From this conditional PDF, the symbol log likelihood ratio (LLR) can be derived easily:
\begin{equation} \label{llr}
\lambda^{(t)}(y) = \ln \frac{\mathrm{Pr} [Y^{(t)} = y | Z^{(t)} = 0]}{\mathrm{Pr} [Y^{(t)} = y | Z^{(t)} = 1]}
= \ln \left[ \cosh \frac{2y}{\sigma^2} \right] - \frac{2}{\sigma^2}.\
\end{equation}
{Given a degraded channel}, we can make the best estimation of 
$Z^{(t)}$ only from $\lambda^{(t)}$.
This LLR expression is a special case of the 
LLR expression derived by Sula et al. \cite{Sula17}.

We return to the argument of the case in which terminals A and B 
employ a binary linear code $C$.
Because of the linearity of the code $C$, it is clear that 
$(Z^{(1)},\ldots, Z^{(n)})$ also belongs to $C$.
From this fact, {maximum likelihood (ML) decoding on the degraded channel} is definable as
\begin{equation} \label{iidml}
(\hat z_1, \ldots, \hat z_n) = \arg \max_{(z_1,\ldots, z_n) \in C}
\prod_{t = 1}^ n L(y_t | z_t), 
\end{equation}
where the likelihood functions are defined as
\begin{equation}
\begin{aligned}
L[y|1] & \triangleq F(y; 0, \sigma^2), \\
L[y|0] & \triangleq \frac{1}{2} F(y; -2, \sigma^2) +\frac{1}{2} F(y; 2, \sigma^2).
\end{aligned}
\end{equation}

This ML rule is sub-optimal because the likelihood is based on the {degraded channel}.
Irrespective of its sub-optimality, the assumption of the {degraded channel} makes
the structure of a decoder simple; it also {enables us to use} known channel 
coding techniques developed for memoryless channels.

Belief propagation (BP) decoding for LDPC codes can be regarded as an 
approximation of ML decoding as a message passing form. It would be 
natural to develop a BP decoding algorithm 
for the binary CAF channel based on the ML rule on the {degraded channel} (\ref{iidml}).
It is not difficult to see that the {\em BP on the degraded channel} coincides with the conventional 
log-domain BP algorithm \cite{Richardson} with symbol LLR expression (\ref{llr}).
A BP decoder of this type has been discussed in \cite{Sula17} \cite{Ullah17}.
A marked advantage of the {BP on the degraded channel}
is that it can be implemented easily based on 
a practical BP decoder for the additive white Gaussian noise (AWGN) channel merely by 
replacing an LLR computation unit.

\section{Density evolution {for degraded channels}}\label{S3}

We use DE to study BP thresholds of 
{the degraded channel}.
In this section, we first introduce population dynamics DE and estimate the BP threshold for uncoupled regular LDPC codes.
Subsequently, the BP threshold for SC-LDPC is then evaluated.

\subsection{Density evolution for asymmetric channels}\label{S3-1}

For simplicity, we specifically examine $(d_l,d_r)$-regular LDPC codes, where $d_l$ and $d_r$ respectively represent 
the variable and check node degrees. Extension to irregular codes is straightforward.
It is noteworthy that we need to handle signal dependent noises~(\ref{vir}) {for the degraded channel}. 
Therefore, we cannot rely on the zero codeword assumption in a DE analysis. 
Here we follow Wang's DE formulation \cite{Wang05} to overcome this difficulty.

The conditional PDF $P^{(l)}(m|z)$ (resp. $Q^{(l)}(\hat{m}|z)$) denote 
the PDF of a message $m$ from a variable node to a check node
(resp. $\hat{m}$ from a check node to a variable node) with transmitted word $z$ at the $l$-th step.
The distribution of LLR of the {degraded} channel is denoted by $P^{(0)}(z)$. 
Those PDFs depend on a transmitted word because of the asymmetric characteristics of the channel. 
Let $\Gamma(P_A)\triangleq P_A\circ \gamma^{-1}$ be a density transformation for a random variable $A$ with distribution $P_A$~\cite{Wang05}
where $\gamma:\mathbb{R}\rightarrow \{0,1\}\times [0,\infty)$, 
$\gamma(m) \! \triangleq\!  \left(1_{m\le 0},\ln\coth\left|{m}/{2}\right|\right)$ 
with an indicator function $1_{\{\,\cdot\,\}}$.

The DE equations for binary asymmetric channels~\cite{Wang05} are given as
\begin{align}
P^{(l)}&(m|z)\!=\!P^{(0)}(z)\otimes \left(Q^{(l-1)}(\hat{m}|z\right)^{\!\otimes(d_l-1)}, \label{eq_d1}\\
Q^{(l)}&(\hat{m}|z)\!=\!\Gamma^{-1}\left(\left\{\Gamma\!\left(\frac{P^{(l)}({m}|0)\!+\!P^{(l)}({m}|1)}{2}\right)\right\}^{\!\otimes(d_r-1)}\right.\nonumber\\
&\!\left.+(-1)^z\!\left\{\Gamma\!\left(\frac{P^{(l)}({m}|0)\!-\!P^{(l)}({m}|1)}{2}\right)\right\}^{\!\otimes(d_r-1)}\right), \label{eq_d2}
\end{align}
where $\otimes$ denotes the convolution operator on PDFs.
Although these convolutions of PDFs can be efficiently evaluated with fast Fourier transformation, 
numerical evaluation entails huge computational costs.
We use an alternative approach, population dynamics \cite{Mezard}, to reduce computational complexity because 
DE analysis for SC-LDPC codes 
 examines a number of DE equations simultaneously.

Equations (\ref{eq_d1}) and~(\ref{eq_d2}) have 
equivalent forms called replica-symmetric cavity equations~\cite{Mezard}, which read
\begin{align}
P^{(l)}(m|z)&\!=\!\int dyL[y|z]\int \prod_{s=1}^{d_l-1}d\hat{m}^{(s)}Q^{(l-1)}(\hat{m}^{(s)}|z)\nonumber\\
&\times\delta\left(m-\lambda(y)-\sum_{s=1}^{d_l-1}\hat{m}^{(s)}\right), \label{eq_d3}\\
Q^{(l)}(\hat{m}|z)&\!=\!\frac{1}{2^{d_r-2}}
\sum_{\{z^{(s)}\}\in S}
\int \prod_{s=1}^{d_r-1}d{m}^{(s)}P^{(l)}({m}^{(s)}|z^{(s)})\nonumber\\
&\times\!\delta\left(\hat{m}\!-\!2\tanh^{-1}\left[\prod_{s=1}^{d_r-1}\tanh\left(\frac{{m}^{(s)}}{2}\right)\right]\right),\! \label{eq_d4}
\end{align}
where $\lambda(y)$ denotes the LLR defined as the r.h.s. of (\ref{llr}) and 
$$S\triangleq\left\{\{z^{(s)}\}\in \{0,1\}^{d_r}:\bigoplus_{s=1}^{d_r}z^{(s)}=0, z^{(d_r)}=z\right\}.$$

In Algorithm 1, we describe a procedure of the population dynamics DE.
In population dynamics, the PDFs $P(\cdot|z)$ and $Q(\cdot|z)$ ($z\in\{0,1\}$) are approximated
to histograms (populations) of $N$ samples denoted by, e.g., $\{\nu_i^0\}$ ($i\in [N] \triangleq \{1,\dots,N\}$).
Parameter $N$ is the population size. 
The DE equations are exactly solved in the large-$N$ limit.
Each sample is updated recursively by an update rule written in a delta function $\delta(\cdot)$ 
in~(\ref{eq_d3}) or~(\ref{eq_d4}).
{After each iteration is completed, we can estimate bit error rate (BER) at the step given as
\begin{equation} \label{ber}
\mathrm{BER}(l) \triangleq \frac{1}{2}P(\hat{z}=1|z=0)+\frac{1}{2}P(\hat{z}=0|z=1),
\end{equation}
where $\hat{z}$ represents a decoded bit via the sign of a message $m_1$ at a variable node.
The distribution of a message $m_1$ is obtained as
\begin{align}
{P}^{(l)}(m_1|z)&\!=\!\int dyL[y|z]\int \prod_{s=1}^{d_l}d\hat{m}^{(s)}Q^{(l-1)}(\hat{m}^{(s)}|z)\nonumber\\
&\times\delta\left(m_1-\lambda(y)-\sum_{s=1}^{d_l}\hat{m}^{(s)}\right). \label{eq_d31}
\end{align}
This distribution is also evaluated similarly to the population dynamics DE.
}
Although the recursion should continue until every population converges,
it stops at the maximum iteration step $T$ in practice.
{We confirm that, for simple AWGN channels, the algorithm with $N\!=\!10^5$ and $T\!=\!2000$
 estimates the BP threshold well.}

\begin{figure}[t]\label{alg}
 \removelatexerror
  \begin{algorithm}[H]
   \caption{Population Dynamics DE.}
  \begin{algorithmic}[1]
   \INPUT Population size $N$, Maximum iteration $T$ 
   \OUTPUT Populations $\{\nu_i^0\}$, $\{\nu_i^1\}$, $\{\hat{\nu}_i^0\}$, and $\{\hat{\nu}_i^1\}$ ($i\in[N]$)
   \State Initialization: $\nu_i^0=\nu_i^1=0$
   \For{$l= 1$ to $T$} 
      \For{$z=0$ to $1$} \Comment{Update of $\{\hat{\nu}_i^z\}$ ($Q(\cdot|z)$)}
        \For{$i= 1$ to $N$}
          \State{Draw $z(1),\dots,z(d_r-1)$ uniformly in $\{0,1\}$ to satisfy $z\oplus \left(\bigoplus_{s=1}^{d_r-1}z(s)\right)=0$.}
      	  \State{Draw $i(1),\dots,i(d_r-1)$ uniformly in $[N]$.}
          \State{$\hat{\nu}_i^z \leftarrow 2\tanh^{-1}\left[\prod_{s=1}^{d_r-1}\tanh\left({\nu}_{i(s)}^{z(s)}/2\right)\right]$.}
        \EndFor
      \EndFor
      \For{$z= 0$ to $1$} \Comment{Update of $\{{\nu}_i^z\}$ ($P(\cdot|z)$)}
        \For{$i= 0$ to $N$} 
 	  \State{Draw $y$ {from} $L[y|z]$.}
      	  \State{Draw $i(1),\dots,i(d_l-1)$ uniformly in $[N]$.}
          \State{${\nu}_i^z \leftarrow \lambda(y)+\sum_{s=1}^{d_l-1}\hat{\nu}_{i(s)}^{z}$.}
        \EndFor
      \EndFor
   \EndFor
  \end{algorithmic}
  \end{algorithm}
\end{figure}

\begin{figure}[t]
\centering
\includegraphics[width=0.95\linewidth]{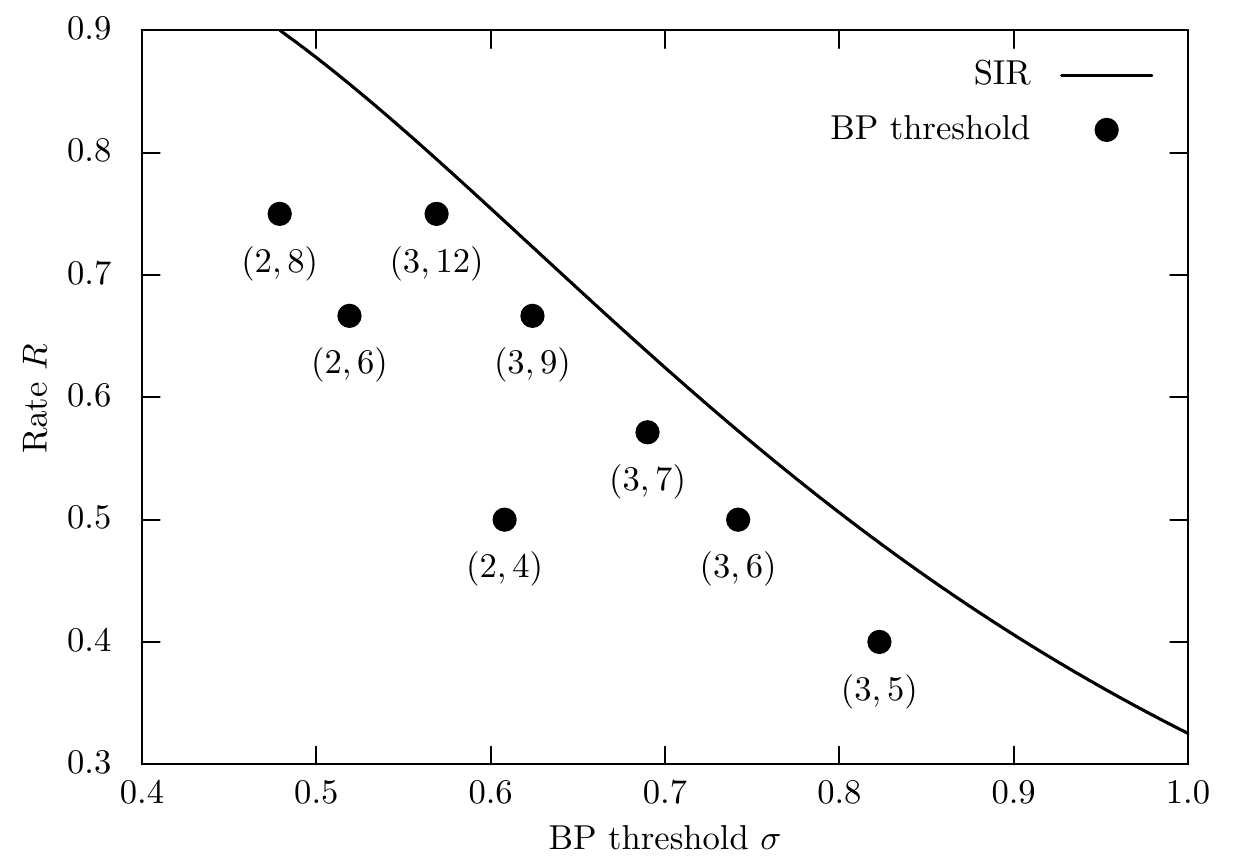}
\caption{BP thresholds for regular LDPC code ensembles over {degraded channels} versus the code rate.
The solid line represents the symmetric information rate of the channel.
}\label{zu_s1}
\end{figure}

We evaluate a BP threshold defined as a threshold of $\sigma$ of the {degraded} channel~(\ref{vir}) below which
 LDPC codes are typically decodable by a BP decoder.
As a MAP threshold, we use the {\em symmetric information rate} (SIR) $\sigma_{sym}(R)$
defined as a solution of 
$C_{sym}(\sigma_{sym}(R))=R$ for code rate $R$, where
\begin{align}
C_{sym}(\sigma) & \triangleq  -\int_{-\infty}^{\infty} P(y) \log_2 P(y) dy \nonumber\\
+&\frac{1}{2} \int_{-\infty}^{\infty} L[y|0] \log_2 L[y|0] dy 
+ \frac{1}{4} \log_2 (2 \pi \sigma^2 e), 
\end{align}
{denotes the symmetric information rate $I(Y;X_1+X_2)$ of {the degraded channel} (\ref{channel_model}) 
and $P(y)  \!=\! (1/2) L[y|0] \!+\! (1/2) L[y|1]$ is the PDF of a received symbol. 
{Here, we omit time index $t$ because of the assumption of {the degraded channel}.}

The BP thresholds of various regular LDPC ensembles versus the code rate are shown in Fig.~\ref{zu_s1}.
We search BP thresholds by evaluating BER using the population dynamics DE with $N\!=\!10^5$ and $T\!=\!2000$. 
The BP thresholds have a gap to the SIR as predicted in~\cite{Sula17}.

\subsection{Spatial coupling coding for {degraded} channels}\label{S3-2}
Next we examine the SC-LDPC codes.
As described in this paper, we examine the simplest $(d_l,d_r,L)$-LDPC codes with chain length $L$
where $k \triangleq d_r/d_l$ and $\hat{d}_l \triangleq (d_l-1)/2$ are integers.
The protograph is then uniquely defined~\cite{Kudekar11}, which makes the structure of the population dynamics DE simple.
The DE analysis for general protograph codes is left as open here.

A protograph of $(d_l,kd_l)$-SC-LDPC codes is represented by $k$ variable nodes and one check node,
e.g., (a) of Fig.~\ref{zu_s2}. 
To construct a protograph of SC-LDPC codes, we prepare $L$ copies of the protograph of an uncoupled code and 
attach $\hat{d}_l$ check nodes to each side of the copy.
Edges of the protograph are then assigned from a variable node
to check nodes within ``distance'' $\hat{d}_l$, e.g., (b) of Fig.~\ref{zu_s2} where $(d_l,d_r,L)=(3,6,5)$.
Consequently, one obtains $L$ bundles of $k$ variable nodes labeled by $i\in [L]$, and
 $L+2\hat{d}_l$ check nodes labeled by $a\in\{-\hat{d}_l+1,\dots,L+\hat{d}_l\}$, where
 check nodes labeled from $1$ to $L$ are derived from original protographs.
The design rate is given as $1-(L+2\hat{d}_l)/(kL)$, which recovers that of uncoupled codes as $L\rightarrow\infty$.
 
\begin{figure}[t]
\centering
\includegraphics[width=0.8\linewidth]{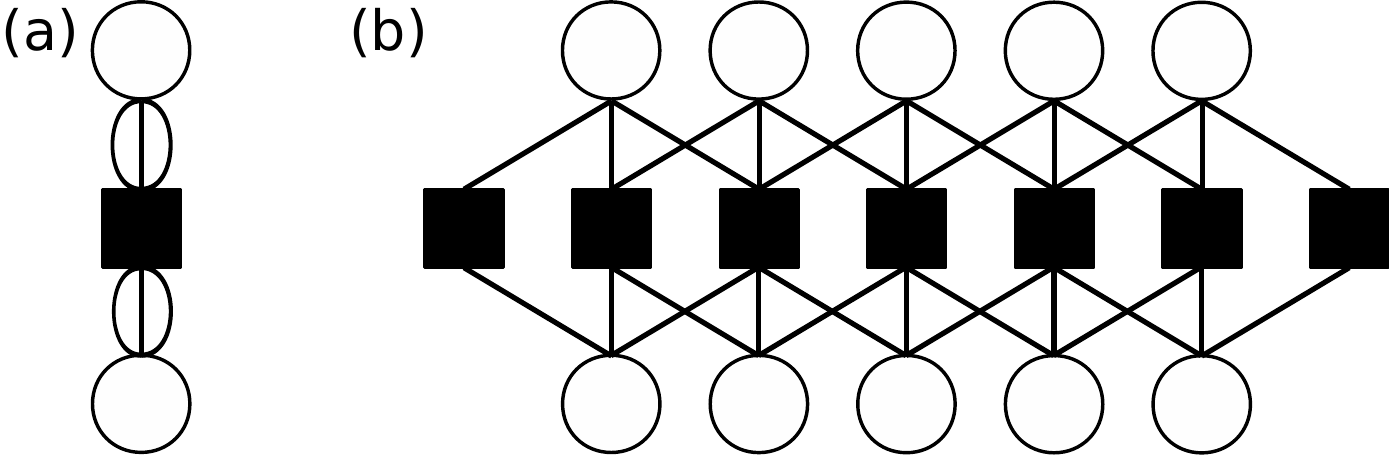}
\caption{Protograph of (a) $(3,6)$-LDPC codes and (b) $(3,6,5)$-SC-LDPC codes.
}\label{zu_s2}
\end{figure}

The BP thresholds can be evaluated for SC-LDPC codes with finite $L$. 
In a protograph, each variable and check node has
a PDF $P^{(l)}(m|z)$ and $Q^{(l)}(\hat{m}|z)$ of messages, as in the last subsection.
Those PDFs are propagated as messages on a protograph.
From the symmetric structure in each bundle,
$P^{(l)}_{i\rightarrow a}(m|z)$ denotes the PDF of message $m$ as a message
 from a variable node in the $i$-th bundle 
 to a check node $a$ at the $l$-th step.
Similarly, let us denote the PDF of message $\hat{m}$ as a message
 from a check node $a$ to a variable node in the $i$-th bundle 
  by $Q^{(l)}_{a\rightarrow i}(\hat{m}|z)$.
DE equations of the {degraded channel} and $(d_l,d_r,L)$-SC-LDPC codes then read 
\begin{align}
&P^{(l)}_{i\rightarrow a}(m|z)=
\int dyL[y|z]\int \prod_{b\in N(i)\backslash a}\!d\hat{m}_bQ^{(l-1)}_{b\rightarrow i}(\hat{m}_b|z)\nonumber\\
&\mathrel{\phantom{P^{(l)}_{i\rightarrow a}(m|z}}\times\delta\left(m-\lambda(y)-\sum_{b\in N(i)\backslash a}\hat{m}_b\right), \label{eq_n3}\\
&Q^{(l)}_{a\rightarrow i}(\hat{m}|z) \!=\!\frac{1}{2^{d_r-2}}\!\sum_{\{z^{(s)}_j\}\in S'}
\int \prod_{s=1}^{k-1}d{m}^{(s)}_i\!P_{i\rightarrow a}^{(l)}({m}^{(s)}_i|z^{(s)}_i)\nonumber\\
&\times \prod_{j\in N(a)\backslash i}\left(\prod_{s=1}^kd{m}^{(s)}_jP_{j\rightarrow a}^{(l)}({m}^{(s)}_j|z^{(s)}_j)\right)\nonumber\\
&\times\delta\left(\hat{m}-2\tanh^{-1}\left[, 
\prod_{(j,s)\neq (i,k)} \tanh\left(\frac{{m}^{(s)}_j}{2}\right)\right]\right), \label{eq_n4}
\end{align}
where $N(\cdot)$ is a set of neighboring nodes in a protograph and
 $$S'\!\triangleq\!\left\{\{z^{(s)}_j\}_{j\in N(a)}^{s\in [k]}\!\in\! \{0,1\}^{d_r}:\bigoplus_{j,s}z^{(s)}_j\!=\!0, z_{i}^{(k)}\!=\!z\right\}.$$
A protograph of uncoupled LDPC codes recovers~(\ref{eq_d3}) and~(\ref{eq_d4}).

The population dynamics {for SC-LDPC codes} is implemented as an extension of Algorithm 1.
In this case, we prepare $4Ld_l$ populations with size $N$ 
to approximate PDFs $P^{(l)}_{i\rightarrow a}(\cdot|z)$ and $Q^{(l)}_{a\rightarrow i}(\cdot|z)$.
Fig.~\ref{zu_s3} shows dynamics of BER of each variable node in $(3,6,25)$-SC-LDPC codes
when $N=10^4$ and $\sigma=0.78$.
It is apparent that they decrease from each side of the chain, as observed in the symmetric channel case~\cite{Kudekar11}.
The BERs vanish after the 169th step, indicating that the code is decodable.

\begin{figure}[t]
\centering
\includegraphics[width=0.95\linewidth]{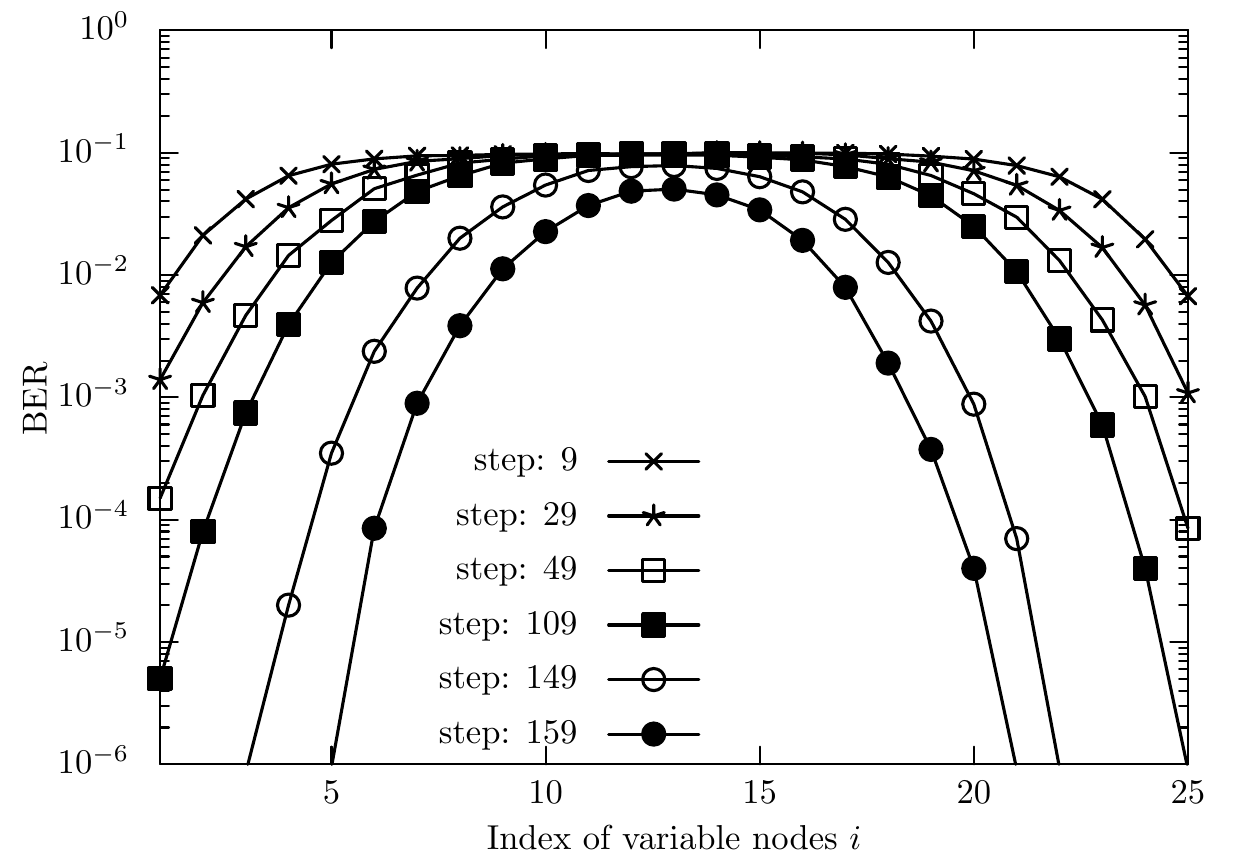}
\caption{BER of each variable node in $(3,6,25)$-SC-LDPC codes with several DE steps evaluated using DE population dynamics for a {degraded channel}.
}\label{zu_s3}
\end{figure}

\begin{figure}[t]
\centering
\includegraphics[width=0.95\linewidth]{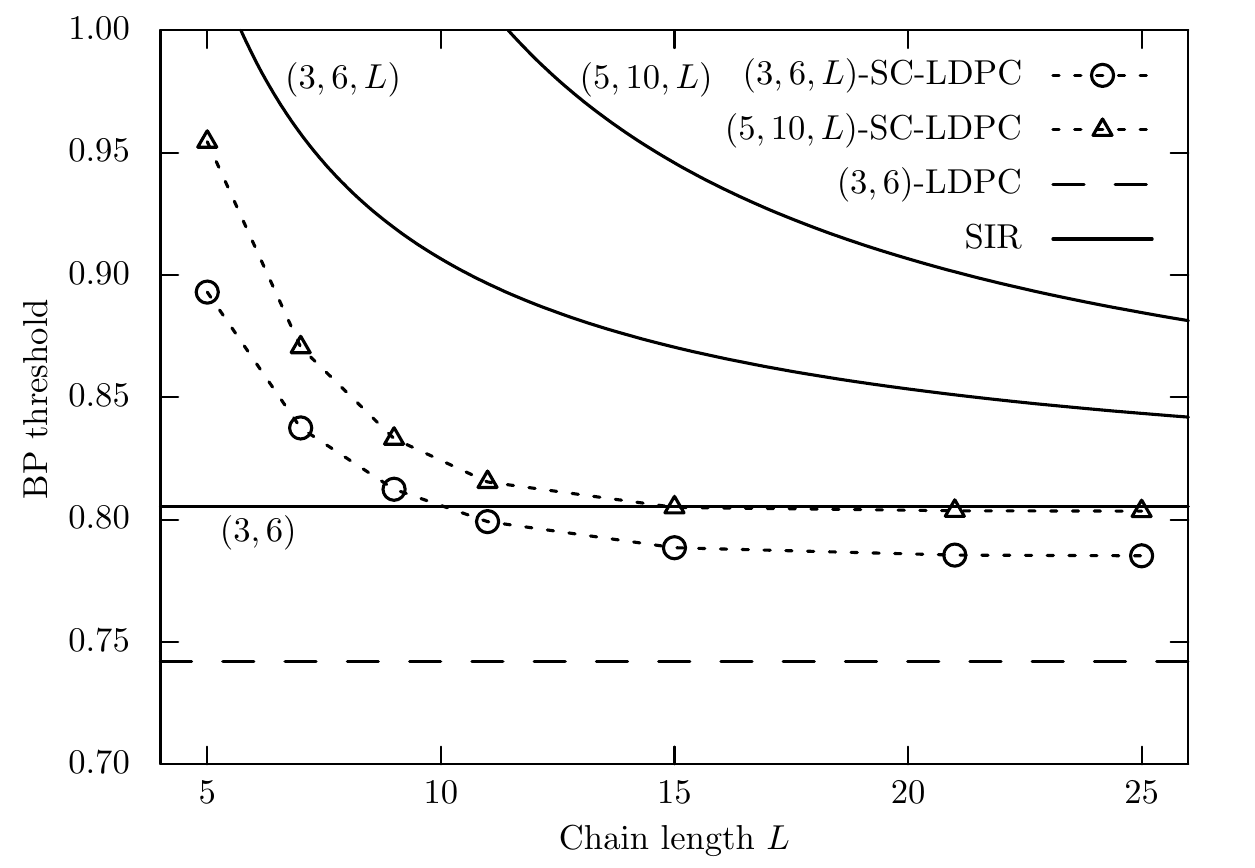}
\caption{{BP thresholds of (SC-)LDPC codes on a {degraded channel}} as a function of chain length $L$.
{Circles and triangles respectively represent the BP thresholds of $(3,6,L)$-SC-LDPC codes and $(5,10,L)$-SC-LDPC codes.}
The long-dashed line represents the BP threshold of uncoupled $(3,6)$-LDPC codes whereas 
each solid line shows the symmetric information rate (SIR) {with the design rate corresponding to a labeled ensemble}.
The SIRs of both SC-LDPC codes converge to that of $(3,6)$-LDPC codes as $L\rightarrow\infty$.
The estimated BP thresholds of $(3,6,L)$-SC-LDPC codes and $(5,10,L)$-SC-LDPC codes in the large-$L$ limit are, respectively, $0.785$ and $0.803$.
}\label{zu_s4}
\end{figure}

{Figure}~\ref{zu_s4} presents the BP threshold of SC-LDPC codes and the {corresponding SIRs}.
In population dynamics, we use $N\!=\!10^5$ and $T\!=\!2000$.
The results indicate that the BP thresholds decrease monotonously as $L$ increases.
{As explained above, the design rate of SC-LDPC codes converges to $1/2$ in the large-$L$ limit.
It is therefore an interesting question whether the decoding performance of the SC-LDPC codes is superior to that of the LDPC codes
as $L\rightarrow\infty$, or not.
To answer the question, BP thresholds by the population dynamics DE are extrapolated by the following function:
\begin{equation}
\sigma_{\mathrm{BP}}=\sigma_{\mathrm{BP}}^\infty+a\exp(-bL^c),
\end{equation}
where $\sigma_{\mathrm{BP}}^\infty$ is the BP thresholds in the large-$L$ limit and 
$a$, $b$, and $c$ are constant parameters.
{As a result,} the limiting values of $(3,6,L)$-SC-LDPC codes and $(5,10,L)$-SC-LDPC codes are estimated respectively as $0.785$ and $0.803$,} 
which lie between the BP threshold $0.742$ of the uncoupled $(3,6)$-LDPC codes 
and the corresponding SIR $0.805$.
The same is true for $(3,9,L)$-SC-LDPC codes: The spatially coupling coding achieves $0.647$ ($L\rightarrow \infty$) 
whereas the BP threshold and the SIR
of uncoupled codes are given respectively by $0.624$ and $0.666$.
It is noteworthy that our evaluation underestimates BP thresholds because $T\!=\!2000$ is generally inadequate. 
A BP decoder for spatially coupled codes is known to need a large number
of iterations before convergence~\cite{Kudekar11}, especially around the threshold.
These facts suggest that the spatial coupling coding improves BP thresholds
although whether it achieves the MAP threshold
 or not remains an open question.

\section{Comparison with SD scheme}\label{S4}
Finally, we compare the CAF scheme with the SD scheme. 
For simplicity, we first consider the case 
in which we use the linear random codes and ML decoding on the degraded channel \eqref{channel_model} in the MAC phase.
In this case, the mutual information rate $I(Y; X_A+X_B)$ based on \eqref{channel_model} is achievable.
{Therefore, as the solid line in Fig. \ref{COM}, we plot the corresponding symmetric information rate as the decodable noise threshold by ML decoding.}
{In addition, as in Sec. \ref{S3},} we have already {evaluated
the asymptotic performance of the $(3,6)$-LDPC codes and $(3,9)$-LDPC codes
and that of the $(3,6,L)$-SC-LDPC codes and $(3,9,L)$-SC-LDPC codes, which are also shown in Fig. \ref{COM}
{and presented in Table~\ref{tab_th}}.
For spatially coupled codes, we show the estimated BP threshold in the large-$L$ limit extrapolated from numerical results with finite $L$,
as described in the last section.}

In the SD scheme, at the MAC phase,
both terminals A and B encode their messages $x_A$ and $x_B$ with the same size
using different linear codes.
Then, the relay R decodes both messages $x_A$ and $x_B$.
In the broadcast phase, relay R sends $x_A$ and $x_B$ to the respective terminals.
The broadcast phase is simple transmission. Therefore,
it has larger capacity than the MAC phase.
For simplicity, under the channel \eqref{channel_model}, we consider the case 
when we employ the random codes and ML decoding on the degraded channel in the MAC phase.
In this case, the mutual information rate $I(Y;X_A,X_B)$ based on \eqref{channel_model}
is achievable \cite{Ahlswede,Liao}.

Here, the rate represents the rate of $x_A$, which is the same as the rate of $x_B$
because the powers from both sides are equal in the channel \eqref{channel_model}.
If the terminals use the same linear codes, relay R cannot decode $x_A$ and $x_B$ for the following reason.
When both message lengths are $\ell$,
relay R cannot distinguish the messages 
$\{(x_A,x_B)\in \mathbb{F}_2^{2 \ell}: x_A+x_B=(1, \ldots, 1)\}$.
However, if they choose their code independently by random coding, then
the mutual information rate is achievable in this scheme.
{Therefore, we plot the corresponding SIR as the decodable noise threshold as the dashed line in Fig. \ref{COM}.}

{To compare the regular LDPC codes and SC-LDPC codes in the CAF scheme},
we specifically examine the numerical results {of the BP thresholds}
in the SD scheme.
Yedla et al.~\cite{Yedla} calculated the asymptotic performance of 
$(3,6)$-LDPC codes and $(3,6,64,5)$-SC-LDPC codes~\footnote{{the numerical estimation of 
BP thresholds are extracted from Fig. 7 in~\cite{Yedla}.}}.
{We show the BP thresholds of these codes in Fig. \ref{COM} and summarize them in Table~\ref{tab_th}.}

These numerical comparisons show that the CAF scheme is advantageous when the rate is {higher} than $1/2$.
This comparison demonstrates that {if the standard deviation is lower than about $0.8$},
the CAF scheme {is expected to exhibit better performance in terms of the SIR}.
{For $R=2/3$, we cannot compare BP thresholds of LDPC codes in two schemes directly because that of the SD scheme is unavailable.
However, even the asymptotic performance of the $(3,9)$-LDPC codes and $(3,9,L)$-SC-LDPC}
of the CAF scheme surpasses the SD scheme with the random coding and {ML decoding}, which has better performance 
than the regular LDPC codes of the same scheme.

For $R=1/2$, both schemes have almost equivalent performance with the random coding and {ML decoding}.
Furthermore, both schemes have almost equivalent performance even with the SC-LDPC codes.
However, the implementation costs of their decoders differ. 
In the CAF scheme, the decoder can be implemented by modifying the conventional {BP} decoder
because it can be regarded as a decoder with an asymmetric channel.
In the SD scheme, the {BP} decoder must reflect the multiple access structure of the encoder,
which increases the decoder complexity~\cite{Yedla}. 
Therefore, we conclude that the CAF scheme is better than the SD scheme when $R=1/2$.

\begin{figure}[t]
    \centering
    \includegraphics[width=0.95\hsize]{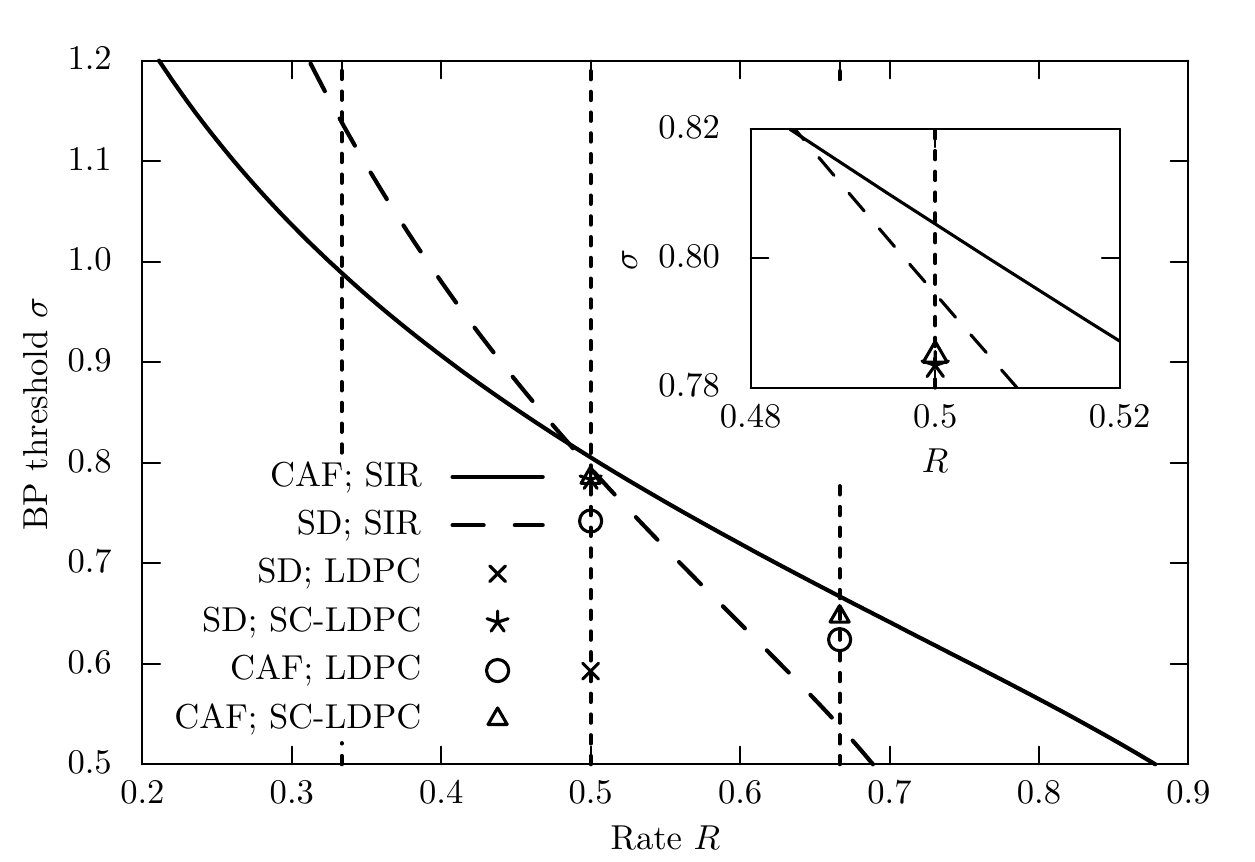}
    \caption{{The BP threshold $\sigma$ as a function of the rate $R$.
The solid line shows the symmetric information rate of the CAF scheme whereas 
the dashed line represents that of the SD scheme.
The dotted vertical lines show rates of $1/3$, $1/2$, and $2/3$.
Cross marks and asterisks respectively denote BP thresholds for $(3,6)$-LDPC codes and $(3,6,64,5)$-LDPC codes
under the SD scheme \cite{Yedla}.
Circles denote BP thresholds for $(3,6)$-LDPC codes ($R=1/2$) and $(3,9)$-LDPC codes $(R=2/3)$ 
under the CAF scheme, whereas triangles represent those of corresponding $(3,6,L)$-SC-LDPC codes and $(3,9,L)$-SC-LDPC codes in the large-$L$ limit.}
}
    \label{COM}
\end{figure}

\begin{table}[t]
\begin{center}
\caption{BP thresholds of (SC-)LDPC codes ($L\rightarrow \infty$) and SIR in the CAF and SD schemes}
\label{tab_th}
  \begin{tabular}{c|c|c|c} \hline \hline
    (SC-)LDPC/SIR & Scheme & Rate & BP threshold/SIR\\ \hline
    $(3,6)$ & CAF & $1/2$ & $0.742$ \\ \hline
    $(3,6,L)$ & CAF & $1/2$ & $0.785$ \\ \hline
    SIR & CAF & $1/2$ & $0.805$ \\ \hline
    $(3,6)$~\cite{Yedla} & SD & $1/2$ & $0.592$ \\ \hline
    $(3,6,64,5)$~\cite{Yedla} & SD & $1/2$ & $0.783$ \\ \hline
    SIR & SD & $1/2$ & $0.794$ \\ \hline\hline
    $(3,9)$ & CAF & $2/3$ & $0.624$ \\ \hline
    $(3,9,L)$ & CAF & $2/3$ & $0.647$ \\ \hline    
    SIR & CAF & $2/3$ & $0.666$ \\ \hline
    SIR & SD & $2/3$ & $0.537$ \\ \hline
  \end{tabular}
\end{center}
\end{table}

\section{Summary}
As described in this paper, asymptotic behavior of LDPC codes and SC-LDPC codes for the CAF relaying are examined.
Combining the population dynamics DE with DE formulas for asymmetric channels, BP thresholds of regular LDPC codes 
are evaluated.
Additionally, we provided the DE equations of $(d_l,d_r,L)$-SC-LDPC codes and performed the population dynamics DE.
Results show that the spatial coupling coding improves the BP thresholds of two-way relay channels.
{We also provide a theoretical demonstration that the CAF scheme potentially exhibits better performance than the SD scheme
in terms of mutual information rate {in the high-rate region}.
Moreover, a BP decoder in the MAC phase of the CAF scheme has a rather simpler structure than that for the SD scheme.
These facts suggest that, under BPSK modulation, the LDPC coding and BP decoding described in this paper practically and theoretically overwhelm the SD scheme.
It is therefore an interesting subject for future work for practical applications to study the LDPC coding and BP decoding with a large constellation in the CAF scheme.
}

\section*{Acknowledgements}
The authors are grateful to Mr. Yuta Ishimatsu for executing some of the numerical analyses.
This work is supported
by a JSPS Grant-in-Aid for Scientific Research (A) Grant Number 17H01280.


\begin{thebibliography}{99}

\bibitem{Cover79}
T. M. Cover and A. El Gamal, 
``Capacity theorems for the relay channel,''
 IEEE Trans. Inf. Theory, vol. 25, no. 5, pp. 572-584, Sep. 1979.

\bibitem{Laneman}
J. N. Laneman, D. N. C.Tse, and G. W. Wornell,
``Cooperative diversity in wireless networks: Efficient protocols and outage behavior,''
 IEEE Trans. Inf. Theory, vol. 50, no. 12, pp. 3062-3080, Dec. 2004.


\bibitem{Nazer11} 
B. Nazer and M. Gastpar,
``Compute-and-forward: harnessing interference through 
structured codes,'' 
IEEE Trans. Inf. Theory, vol. 57, no. 10, pp. 6463-6486, Oct. 2011.


\bibitem{Katti08}
S. Katti, H. Rahul, W. Hu, D. Katabi, M. Medard, and J. Crowcroft,
``XORs in the air: practical wireless network coding,''
IEEE/ACM Trans. Networking, vol. 16, no. 3, pp. 497-510, Jun. 2008.

\bibitem{Zhang09}
S. Zhang, and S.-C. Liew,
``Channel coding and decoding in a relay system operated with physical-layer network coding,''
IEEE J. Select. Areas in Commun., vol. 27, no. 5, pp. 788-796, Jun. 2009.

\bibitem{Sula17}
E. Sula, J. Zhu, A. Pastore, S. H. Lim, and M. Gastpar, 
``Compute-forward multiple access (CFMA) with nested LDPC codes,''
\textit{Proc. IEEE Int. Symp. Inf. Theory}, Aachen, Jun. 2017, pp. 2935-2939.

\bibitem{Ullah17}
S. S. Ullah, G. Liva, and S. C. Liew,
``Physical-layer network coding: a random coding error exponent perspective,''
\textit{Proc. IEEE Inf. Theory Workshop}, Kaohsiung, Nov. 2017.

\bibitem{Narayanan}
K. Narayanan, M. P. Wilson, and A. Sprintson, 
``Joint physical layer coding and network coding for bi-directional relaying,''
\textit{Proc. 45th Ann. Allerton Conf. Commun., Contr. Comput.}, Monticello, IL, Sep. 2007, pp. 5641-5654.

\bibitem{Gallager63}
R. G. Gallager, 
\textit{Low-Density Parity-Check Codes},
MIT Press, 1963.


\bibitem{MacKay99}
D. J. C. MacKay, 
``Good error correcting codes based on very sparse matrices,''
IEEE Trans. Inf. Theory, vol. 45, no. 2, pp. 399-431, Mar. 1999.


\bibitem{Richardson}
T. Richardson and R. Urbanke, {\it Modern Coding Theory}, Cambridge University Press, 2008.


\bibitem{Mezard}
M. M\'ezard and A. Montanari, 
\textit{Information, Physics, and Computation},
Oxford University press, 2009.



\bibitem{Wang05}
C.-C. Wang, S. R. Kulkarni, and H. V. Poor,
``Density evolution for asymmetric memoryless channels,''
IEEE Trans. Inf. Theory, vol. 51, no. 12, pp. 4216-4236, Dec. 2005.

\bibitem{Zigangirov99}
A. J. Felstrom and K. S. Zigangirov, 
``Time-varying periodic convolutional codes with low-density parity-check matrix,''
IEEE Trans. Inf. Theory, vol. 45, no. 5, pp. 2181-2190, Mar. 1999.

\bibitem{Lentmaier10}
M. Lentmaier, A. Sridharan, K. S. Zigangirov, and D. J. Costello, Jr., 
``Iterative decoding threshold analysis for LDPC convolutional codes,''
IEEE Trans. Inf. Theory, vol. 56, no. 10, pp. 5274-5289, Oct. 2010.

\bibitem{Kudekar13}
S. Kudekar, T. Richardson, and R. L. Urbanke,
``Spatially coupled ensembles universally achieve capacity under belief propagation,''
IEEE Trans. Inf. Theory, vol. 59, no. 12, pp. 7761-7813, Dec. 2013.

\bibitem{Hern} B. Hern and K. Narayanan, 
``Joint compute and forward for the two-way relay channel with spatially coupled LDPC codes,''
\textit{2012 IEEE Global Commun. Conf.}, Anaheim, CA, 2012, pp. 2340-2345.


\bibitem{Thorpe04}
J. Thorpe, K. Andrews, and S. Dolinar, 
``Methodologies for designing LDPC codes using protographs and circulants,''
\textit{Proc. IEEE Int. Symp. Inf. Theory}, Chicago, Jun. 2004, p. 238.

\bibitem{Hassani}
S. H. Hassani, N. Macris, and R. Urbanke, 
``Threshold saturation in spatially coupled constraint satisfaction problems,'' 
Journal of Statistical Physics, vol. 150, no.5, pp. 807-850, Dec. 2013.

\bibitem{Yedla09} A. Yedla, H. D. Pfister and K. R. Narayanan, 
``Can iterative decoding for erasure correlated sources be universal?'' 
\textit{2009 47th Annual Allerton Conf. Comm., Control, Comp.}, Monticello, IL, 2009, pp. 408-415.

\bibitem{Nazer07} B. Nazer and M. Gastpar, 
``Computation Over Multiple-Access Channels,'' IEEE Trans. Inf. Theory, vol. 53, no. 10, pp. 3498-3516, Oct. 2007.

\bibitem{Yedla}
A. Yedla, P. S. Nguyen, H. D. Pfister, and K. R. Narayanan
``Universal codes for the Gaussian MAC via spatial coupling,''
\textit{2011 49th Annual Allerton Conf. Comm., Control, Comp.}, Monticello, IL, 2011, pp. 1801-1808.

\bibitem{Kudekar11} S. Kudekar, T. J. Richardson, and R. L. Urbanke, 
``Threshold saturation via spatial coupling: Why convolutional LDPC ensembles perform so well over the BEC,''
 IEEE Trans. Inf. Theory, vol. 57, no. 2, pp. 803-834, Feb. 2011.

\bibitem{Ahlswede}
R. Ahlswede, ``Multi-way communication channels,'' 
in Proc. Second Int. Symp. Inf. Theory (Thakadsor, Armenian SSR, Sep. 1971). Budapest,
Hungary: Academia Kiado, 1971, pp. 23-52.

\bibitem{Liao}
H. Liao, ``Multiple access channels,'' Ph.D. dissertation, Dept. Electr. Eng., University of Hawaii, Honolulu, 1972.


\end{thebibliography}
\end{document}